\documentclass[conference]{IEEEtran}

\usepackage{cite}

   \usepackage[pdftex]{graphicx}
   \DeclareGraphicsExtensions{.pdf,.jpeg,.png,.JPG,.jpg}
\usepackage{marginnote}

\usepackage{eso-pic}
\AddToShipoutPicture{
	\raisebox{2\baselineskip}{\makebox[\paperwidth]{\begin{minipage}[t][0.5cm][b]{21cm}\centering\copyright \scriptsize 2019 IEEE. Personal use of this material is permitted. Permission from IEEE must be obtained for all other uses, in any current or future media, including reprinting/republishing this material for advertising or promotional purposes, creating new collective works, for resale or redistribution to servers or lists, or reuse of any copyrighted component of this work in other works.\end{minipage}
}}}

\begin{document}
%
\title{Fabrication of optical components with nm- to mm-scale critical features using three-dimensional direct laser writing}

\author{\IEEEauthorblockN{Yanzeng Li,$^{1}$ Serang Park,$^{1}$ Michael McLamb,$^{1}$ Marc Lata,$^{1}$ Darrell Childers,$^{2}$ and Tino Hofmann$^{1}$}
	\IEEEauthorblockA{$^{1}$\ Department of Physics and Optical Science, University of North Carolina at Charlotte\\
		Charlotte, North Carolina, 28223\\
		$^{2}$\ US Conec,1138 25th Street Southeast\\ Hickory, NC 28602, USA\\
		Email: yli91@uncc.edu}}


%


\maketitle

\begin{abstract}
A powerful fabrication strategy based on three-dimensional direct laser writing for the rapid prototyping of opto-mechanical components with critical features ranging from several hundred nm to a few mm is demonstrated here. As an example, a simple optical fiber connector with optical and mechanical guides as well as integrated micro-optical elements with nano-structured surfaces is designed and fabricated. In contrast to established three-dimensional direct laser writing, two different polymers are combined in the fabrication process in order to achieve a drastic reduction in fabrication time by substantially reducing the ``optical tool path''. A good agreement between the as-fabricated connector and nominal dimensions has been obtained. The developed approach allows the rapid prototyping of optomechanical components with multi-scale critical features. It is, therefore, envisioned to substantially accelerate the development cycle by integrating functional mechanical and optical elements in a single component.  

\end{abstract}


%
\IEEEpeerreviewmaketitle

\section{Introduction}
Three-dimensional direct laser writing (3D-DLW) based on two-photon polymerization is a widely employed additive manufacturing technique since Deubel \textit{et al.} used this approach in 2004 to synthesize infrared photonic crystals \cite{M.Deubel2004direct}.
Since then, this unique additive fabrication technique has been successfully used for the fabrication of structured optical components with $\mu$m- and even nm-scale critical dimensions. 3D-DLW allows the polymerization of a photosensitive material within the free-space volume, a so called voxel, which is substantially smaller compared to voxel sizes which can be achieved using conventional single-photon polymerization techniques. The voxel dimensions 
are typically in the range from hundreds of nm to several $\mu$m. This technique is therefore suitable for the applications ranging from the rapid prototyping of micro-optical components to the synthesis of complex optical metamaterials with sub-wavelength-sized building blocks for the infrared and even visible spectral range \cite{sakellari20173d,moughames2016wavelength,caligiuri2016dielectric}. In recent years the rapid prototyping of optical components applied in various optical fields, including functional coatings \cite{zinkiewicz2015highly,kowalczyk2014microstructured}, metasurfaces \cite{xiong2013structured,jose2019tunable}, structured photonics \cite{li2018high,MarichySR6_2016}, and micro-optics \cite{Gissibl2016two,schmid2018three,Y.Li2017NewMicroLenses} has been shown.

In addition to applications of 3D-DLW for the fabrication of optical elements with microscopic dimensions, recent decades have also witnessed intensive research efforts devoted to the development of mechanical components with macroscopic dimensions (mm-scale) by using the 3D-DLW approach \cite{rodriguez20173d,buckmann2014elasto,tielen2014three}. In comparison with the other techniques for the fabrication of macroscopic mechanical parts, 3D-DLW can fabricate a mechanical part with a complex structural configuration in a single-step without involving additional manufacturing procedures. Furthermore, 3D-DLW simplifies the manufacturing complexity which can often result in a lower manufacturing time and a reduction of the fabrication costs. From an application perspective in an industrial environment, 3D-DLW is a promising tool for rapid prototyping of novel opto-mechanical components and could accelerate the product development cycle.

The capabilities of the 3D-DLW process to fabricate structures with either microscopic or macroscopic features have been extensively studied individually. However, approaches on how to utilize the 3D-DLW technique for the synthesis of structures which contain critical features ranging from nm- to mm-scales, has not been reported yet. 

The ability to rapidly prototype such structures has numerous applications for integrated optics and opto-mechanical elements. In order to obtain high quality components with nm- to mm-scale features in a manageable fabrication time, the use of multiple photosensitive materials and objectives is very helpful as will be shown here. 

In this paper, hetero-3D-DLW for the fabrication of components with nm- to mm-scale critical features is introduced. This novel approach makes use of two photo-resists in order to achieve a substantial reduction in the ``optical tool path'', thereby dramatically reducing the fabrication time. As an example, the fabrication of a optical fiber connector (mm-scale) is described. The connector contains mechanical alignment guides as well as multiple fiber guides which are equipped with microlenses ($\mu$m-scale). In order to further demonstrate the capability to fabricate structures with nm-scale dimensions, the surface of one of the microlenses was coated with an anti-reflective structured surface (ARSS, nm-scale). 




\section{Experiment}
\subsection{Sample design}
\begin{figure}[htb]
	\centering
	\includegraphics[width=0.7\columnwidth,trim=0 50 0 40,clip]{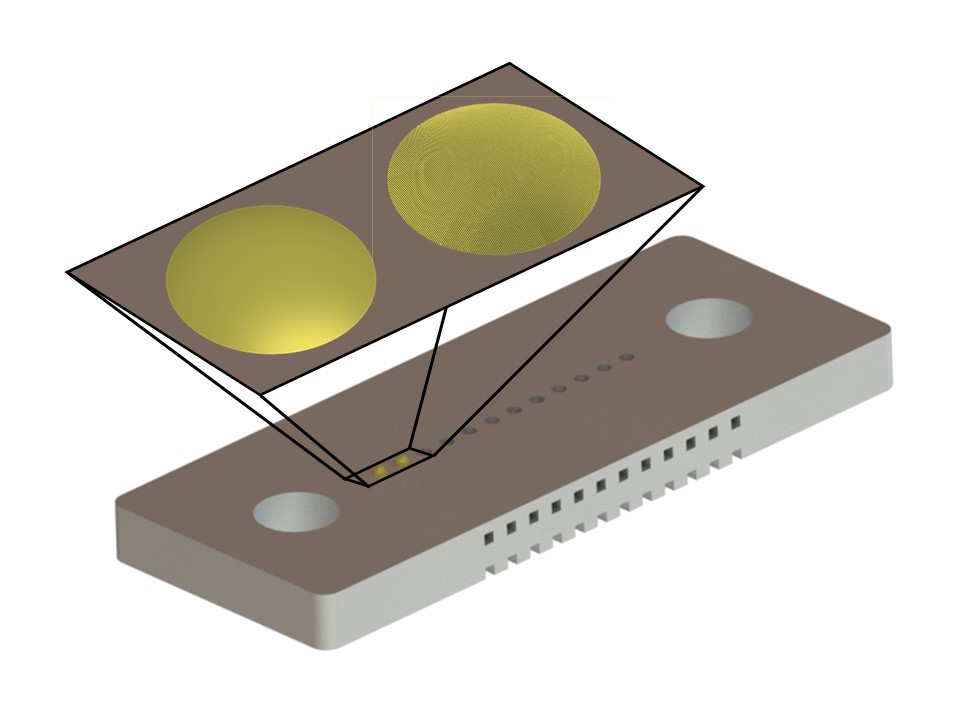}
	\caption{CAD rendering of the fiber connector. The outer dimensions of the connector are on the mm-scale, while it contains elements with drastically smaller critical dimensions. Located at the left and right of the connector are mechanical guides which are designed to guide a receptacle with high-precision onto the connector. Located within the body of the connector are 12 fiber guides. For demonstration purposes 10 of these fiber guides are bare, 2 are equipped with plano-convex microlenses. One of the microlenses is conformally coated with a structured surface to form a anti-reflection coating (the right microlens in the inset).}
	\label{fig:CAD}	
\end{figure}

The necessary procedures required for the 3-DLW-based fabrication of components with nm- to mm-scale critical features are demonstrated here, using a simple fiber connector as an example. The fiber connector was designed to include mm-scale external dimensions. 
It consists of mechanical as well as fiber guides and microlenses with $\mu$m-scale features. The capability to fabricate nm-scale structures is shown by integrating an ARSS which conformally coats the convex surface of one of the microlenses. The fabrication strategies developed here are not limited to this simple example, but can be employed for a wide range of applications where optical components consisting of features with nm- to mm-scale critical features are desired. A CAD rendering of the fiber connector is shown in Fig.~\ref{fig:CAD}.
The fiber connector has the shape of a rectangular cuboid (length=6.4~mm, width=2.5~mm, and height=0.5~mm) and contains 12 cylindrical fiber guides with a diameter of 127~$\mu$m. The cylinder axis of the fiber guides is oriented normal to the fiber connector face. In addition to the fiber guides, the connector also includes two cylindrical mechanical alignment guides with a diameter of 700~$\mu$m. These alignment guides are designed to receive the mechanical alignment pins of a receptacle, thereby ensuring the accurate alignment of the connector and the receptacle. Consequently, this will provide an accurate alignment of the optical fibers within the connector and the receptacle component. 

In order to ensure that the cylindrical fiber guides are accurately fabricated, 12 pairs of micro-channels were included at the bottom of the fiber connector as shown in Fig.~\ref{fig:CAD}. These micro-channels with a cross section of 100~$\mu$m$\times$100~$\mu$m are connected to the fiber guides and are oriented normal to the side planes of the fiber connector. Thereby the micro-channels aid in the removal of unpolymerized monomer after the polymerization step, during the development procedure. This simple micro-fluidic design substantially reduces the development time and ensures the complete removal of monomer within the fiber guides. The micro-fluidic channels further enable the inclusion of microlenses located at the top of the fiber guides. Without the micro-fluidic channels excess monomer would be trapped within the fiber guides between the substrate surface and lenslet located at opposing ends of the fiber guide. 

\begin{figure}[!h]
	\centering
	\includegraphics[width=0.85\columnwidth,trim=0 40 0 40,clip]{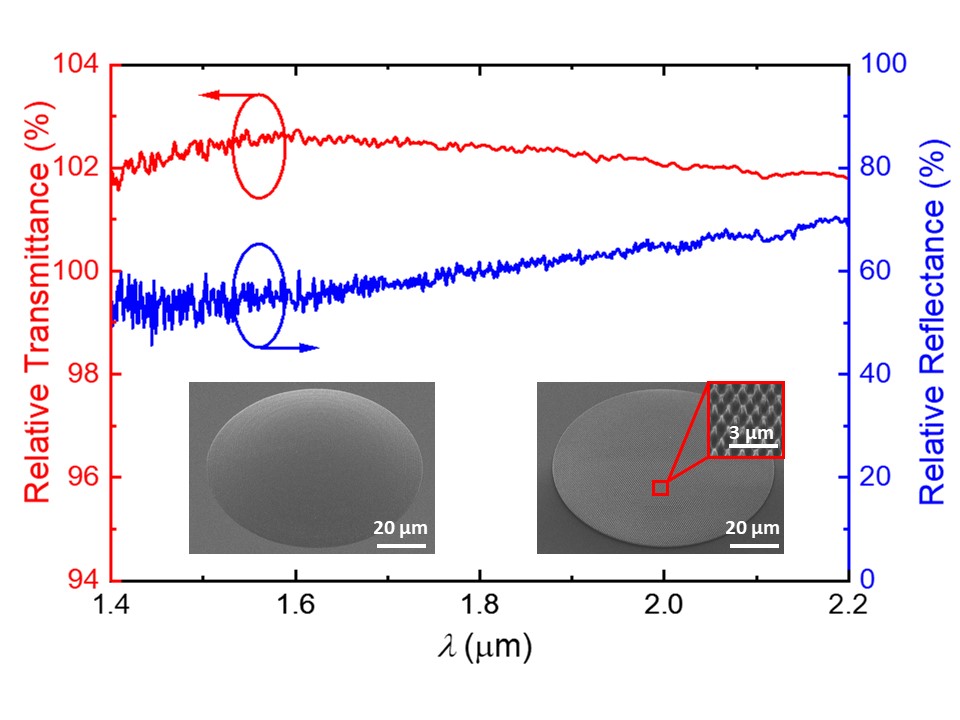}
	\caption{Relative reflectance (blue-solid line) and transmittance
		(red-solid line) spectra of an ARSS-treated microlens (right inset) measured by FTIR microscopy in the range from 1.4 to 2.2~$\mu$m using the absolute reflectance and transmittance of a bare microlens (left inset) as a reference, respectively.}
	\label{fig:ARSS-SEM}
\end{figure}

For demonstration purposes, a pair of microlenses with and without an ARSS coating were included at the top of two fiber guides (as shown in the inset in Fig.~\ref{fig:CAD}). ARSSs are well-known bio-inspired coatings which reduce Fresnel reflection loss \cite{Y.Li2018BroadbandAntireflection}. Here the ARSS are composed of subwavelength conicoid structures arranged in a hexagonal lattice pattern. 
The structural parameters of the conicoid structures, including height, pitch, lateral fill fraction, were optimized for a broadband anti-reflective behavior over spectral range from 1.4~$\mu$m to 2.2~$\mu$m, while the optimal performance (40\% reduction in reflection and 2\% improvement in transmission) occurred at the telecommunication wavelength of 1.55~$\mu$m, as shown in Fig.~\ref{fig:ARSS-SEM}.

The structural features of the ARSS were integrated into the convex surface of a simple planoconvex microlens, with a based diameter of 130 $\mu$m and a curvature of 0.01~$\mu$m$^{-1}$. The effective focal length of the microlens is 189~$\mu$m resulting in an $f$-number of 1.89 for operation at a wavelength of 1.55~$\mu$m. The two insets in Fig.~\ref{fig:ARSS-SEM} show the SEM micrographs of the investigated bare microlens and ARSS coated microlens, respectively. For information about the design of the conformal ARSS coated microlens, the interested reader is referred to our previous publications \cite{Y.Li2018BroadbandAntireflection,li2019near}.

%





\subsection{Sample fabrication}
\begin{figure}[htb]
	\centering
	\includegraphics[width=2.6in]{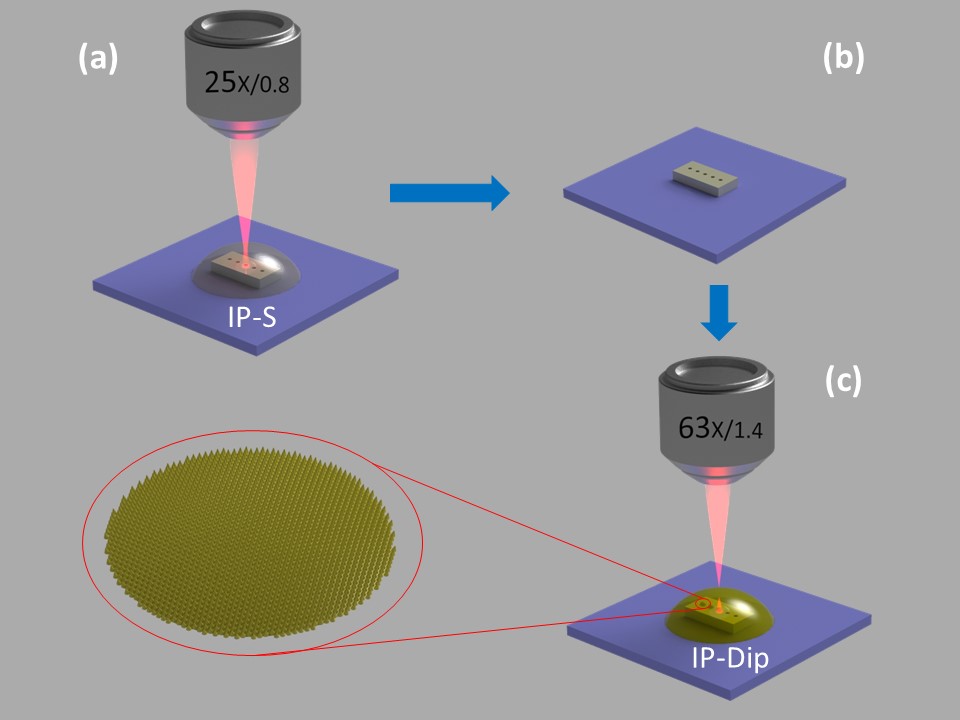}
	\caption{Schematic of the fabrication process. (a) illustrates the polymerization of the components which required the lowest spatial resolution. For this purpose a 25$\times$ immersion objective was used in combination with IP-S photoresist. (b) after the polymerization the access monomer was removed as described in the text. (c) illustrates the subsequent fabrication of the bare and ARSS coated microlenses. The critical features of microlenses and coatings require resolutions in the $\mu$m to nm range. A 63$\times$ immersion objective was used in combination with IP-Dip photoresist in order to achieve this resolution. Note, the objective focal lengths, as shown in Fig.~\ref{fig:schematic}~(a) and~(c), are exaggerated for illustration purposes.}
	\label{fig:schematic}
\end{figure}

The fiber connector including the microlenses with and without the ARSS coating was fabricated using a commercially available 3D-DLW system (Photonic Professional GT, Nanoscribe, GmbH). The employed process steps are illustrated in Fig.~\ref{fig:schematic}. Note, the objective focal lengths, as shown in Fig.~\ref{fig:schematic}~(a) and~(c), are exaggerated for illustration purposes. Both objectives are immersed in the photoresist during the polymerization process. 

As the first step, the features of the fiber connector which require the lowest spatial resolution were fabricated on a glass substrate as shown in Fig.~\ref{fig:schematic}~(a). During this step the rectangular cuboid (see Fig.~\ref{fig:CAD}) including the mechanical and optical guides as well as the micro-fluidic channels is polymerized. In order to expedite the fabrication process a 25$\times$ immersion objective was used in combination with the photoresist IP-S. Due to the high viscosity of the photoresist IP-S combined with the large focal spot of the 25$\times$ objective, voxel dimensions on the order of several micrometers are achievable. This allows for a larger distance between the layers which have to be polymerized as well. This reduction in ``optical tool path'' length substantially reduces the fabrication time while maintaining 
sufficient spatial resolution.


After the first polymerization step illustrated in Fig.~\ref{fig:schematic}~(a), a developing process was applied to remove the unpolymerized IP-S in order to obtain a dry and clean fiber connector, as shown in Fig.~\ref{fig:schematic}~(b). The unpolymerized IP-S, was removed by immersing the sample in propylene glycol monomethyl ether acetate (PGMEA) and, subsequently, in 99.99\% isopropyl alcohol (IPA) for 40~min and 20~min, respectively. After evaporating the excess IPA in dry nitrogen, the fiber connector was subjected to a second polymerization step as shown in Fig.~\ref{fig:schematic}~(c). 

The final step aims at fabricating the microlenses with $\mu$m-scale dimensions and nm-scale features on top of the fiber guiding holes of the fabricated fiber connector, as shown in Fig.~\ref{fig:schematic}~(c). In this process step, a 63$\times$ immersion objective is used in combination with the less viscous IP-Dip. This combination of objective and photo-resist reduces the effective voxel size to a few hundred nanometers and thereby provides sufficient spatial resolution to fabricate both the bare microlens as well as the microlens with the conformal ARSS coating, which consists of nm-scale conicoid features.


After re-immersing the fiber connector inside a drop of the photo-resist IP-Dip, the sample stage was manually controlled to locate the interface of the fiber guides of the fiber connector. Following this manual alignment procedure, a pair of microlenses was fabricated on top of two fiber guides. While one of the microlenses was fabricated with a bare convex surface, a second microlens was fabricated with a conformal ARSS coating.
Following the polymerization, the development process was repeated as before to obtain the final fiber connector consisting of a cuboid connector body, mechanical and fiber guides as well as integrated  microlenses with and without ARSS coating. A top-view optical micrograph is depicted in Fig.~\ref{fig:optical_micrograph} and will be discussed in the next section. It is important to note that the hetero-3D-DLW fabrication approach described above, which combines two different photo-resists and immersion objectives, dramatically reduces the fabrication time. Compared with traditional homo-3D-DLW fabrication, one order of magnitude in fabrication time is saved. The total fabrication time of the fiber connector was approximately 13~hours.

\subsection{Result and discussion}
\begin{figure}[htb]
	\centering
	\includegraphics[width=3in]{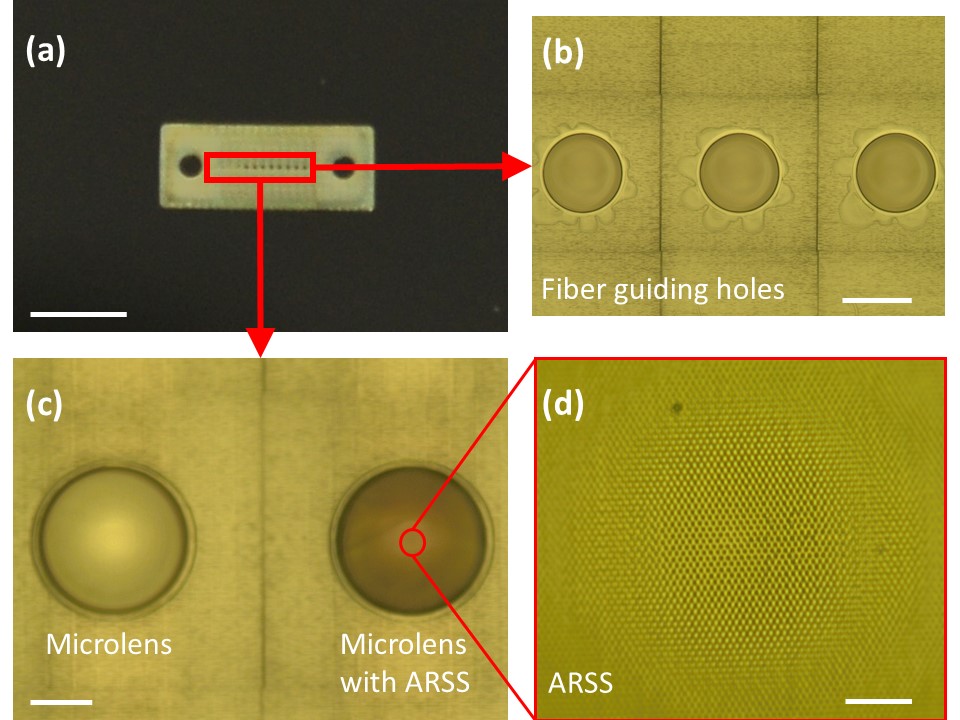}
	\caption{Top-view optical micrographs of the fabricated fiber connector showing the entire connector (a), a detailed micrograph of the fiber guides (b), a (c) comparison of a microlens without (left) and with (right) a conformal ARSS coating, and (d) a detailed micrograph of ARSS surface only. The scale bars in (a), (b), (c), and (d) are 2.5~mm, 100~$\mu$m, 50~$\mu$m, and 10~$\mu$m, respectively.}
	\label{fig:optical_micrograph}	
\end{figure}

The fabricated fiber connector was investigated by using optical microscopy, as shown in Fig.~\ref{fig:optical_micrograph}. In Fig.~\ref{fig:optical_micrograph} (a), the complete cuboid fiber connector is depicted. A row containing 12 fiber guides can be seen in the center of the cuboid between the two larger mechanical guides on each side. The first two fiber guides from the left were selected to be covered by two microlenses to demonstrate the hetero-3D-DLW approach here. The remaining 10 fiber guides were not covered to allow a direct inspection. Fig.~\ref{fig:optical_micrograph} (b) shows a detailed top-view of the fiber guides where no microlenses were added. The diameter of the fiber guides is $127\pm 0.7~\mu$m and in excellent agreement with the design. 

The dark orthogonal lines are the result of multiple exposure due to the stitching of areas, which are polymerized in a single step using the galvanometric scanner of the 3D-DLW system. The galvanometric scanner is tuned to produce a slight overlap of adjacent areas. While this overlap ensures the continuity of the fiber connector, it also results in an additional laser exposure of these overlap areas. This leads to an observable optical contrast. The effects of the over exposure on the infrared optical properties of the employed polymers are not studied. We therefore restricted these overlap areas to sections which have no optical and only mechanical functionality as surface finish and mechanical properties of the polymers are thought to be not impaired. 
 
A top-view micrograph of a bare microlens and an ARSS coated microlens are shown in Fig.~\ref{fig:optical_micrograph}~(c) for comparison. The substantial contrast between the bare microlens shown on the left side of the micrograph (bright) and microlens with ARSS coating shown on the right side (dark) originates from scattering at visible wavelengths. The conicoid constituents of the ARSS are designed to operate in the infrared spectral range. At visible wavelengths, however, the conicoid constituents are larger than the wavelength and hence scatter light significantly. The brightness of the bare microlens further indicates the smoothness of the lens surface that has been well constructed without defects. Figure~\ref{fig:optical_micrograph}~(d) provides a detailed top-view of the ARSS coated microlens to show the structure of the AR constituents, which are homogeneously and conformally distributed in a hexagonal lattice pattern across the convex lens surface. A good agreement on the structural features between the fabricated and designed opto-mechanical component has been obtained. 

\vspace{-0.9cm}
\section{Conclusion}
Hetero-3D-DLW, which has been introduced here as a 3D-DLW process involving multiple polymers, allows the fabrication of opto-mechanical components with multi-scale critical features ranging from the nm- to the mm-scale. The use of different polymers
results in a dramatic reduction of the process time of hetero-3D-DLW compared to approaches where only a single polymer-objective combination is used. This is mainly due to the reduction in the ``optical tool path''. The polymerization of IP-S with a 25$\times$ immersion objective and IP-Dip with a 63$\times$ immersion objective, has been found to allow for the efficient fabrication of mm-scale structures with low spatial resolution requirement (few micrometers) and $\mu$m-scale structures with high spatial resolution requirement (few hundred nanometers). A simple fiber connector with a pair of integrated microlenses with and without ARSS coating has been used to 
illustrate this fabrication approach. A good agreement between the as-printed sample and nominal dimensions was achieved. In contrast to conventional 3D-DLW where only a single polymer objective combination is used, the hetero-3D-DLW approach shown here allows the fabrication of extremely complex optical components. The combination of different polymers and objectives reduces the process time dramatically. We envision hetero-3D-DLW for the rapid prototyping of opto-mechanical components, which could substantially accelerate the development cycle of opto-mechanical devices by integrating mechanical and optical functionality.








\vspace{-0.35cm}
\section*{Acknowledgment}
\noindent The authors are grateful for support from the National Science Foundation (1624572) within the I/UCRC Center for Metamaterials, the Swedish Agency for Innovation Systems (2014-04712), and the Department of Physics and Optical Science of the University of North Carolina at Charlotte.



%


\end{document}